\documentclass{PoS}

\newlength{\halffigwidth}
\usepackage{graphicx}
\usepackage{psfrag}
\usepackage{subfigure}

\title{Gauge fixing methods and Gribov copies effects in lattice QCD}

\ShortTitle{Gauge fixing methods and Gribov copies effects in lattice QCD}

\author{\speaker{Paulo J. Silva} and Orlando Oliveira\\
        Centro de F\'{\i}sica Computacional,
        Departamento de F\'{\i}sica,
        Universidade de Coimbra, P-3004-516 Coimbra,
        Portugal\\
        E-mail: \email{psilva@teor.fis.uc.pt},
        \email{orlando@teor.fis.uc.pt}}

\abstract{We compare two Landau gauge fixing methods, aiming to find the global maximum of the gauge fixing functional. Moreover, a systematic effect of Gribov copies in the gluon and ghost propagators computed in Landau gauge is presented and discussed.}

\FullConference{The XXV International Symposium on Lattice Field Theory\\
		 July 30 - August 4 2007\\
		 Regensburg, Germany}

\begin{document}

\section{Comparing Landau gauge f\mbox{}ixing methods}

The formulation of QCD on the lattice does not require gauge f\mbox{}ixing. However, for example, if one wishes to study the propagators of the fundamental f\mbox{}ields, one must choose a given gauge. A possible choice is the so called Landau gauge, $\partial_{\mu} A_{\mu} =0$. On the lattice, the Landau gauge is implemented by searching for stationary points of 
\begin{equation}
F_{U} [g]  =   \sum_{x,\mu} \mbox{Re} \{ \mbox{Tr} [ g(x) U_{\mu} (x) g^{\dagger }(x+\hat{\mu})]\}
\label{f}
\end{equation}
over the gauge orbit of $U_\mu$. As any other local continuous gauge f\mbox{}ixing condition \cite{singer,killingback}, Landau gauge suffers from the so called Gribov copies problem \cite{gribov}, i.e. there are multiple solutions for the gauge condition in each gauge orbit. This rises the question on the possibility of a non-perturbative def\mbox{}inition of the Landau gauge. 
A solution to this problem is to identify Landau gauge as the search of the unique \cite{GAnt91}, up to a global gauge transformation, absolute maxima of (\ref{f}).
Unfortunately, the search of a global maximum of a function on a multidimensional manifold is far from being trivial.  For the particular case of the gauge under discussion, some methods have been devised which, hopefully, will be able to compute the absolute maximum of $F_U [g]$. 

In this work we compare two of them, namely the CEASD method, described in \cite{cpc2004}, and the smeared gauge f\mbox{}ixing, described in \cite{fat}.
The CEASD method combines an evolutionary algorithm with the steepest descent (SD) method \cite{Davies}. On the other hand, smeared gauge f\mbox{}ixing relies on the smooting of the gauge f\mbox{}ixing hypersurface by smearing the conf\mbox{}iguration.

We tested these two methods on SU(3) $16^4$ pure gauge Wilson action conf\mbox{}igurations \footnote{All the conf\mbox{}igurations used in this work were generated with MILC code \cite{milc}.} for $\beta \in \{ 5.7,  5.8,  5.9,  6.0, 6.2 \}$. For each $\beta$ value, f\mbox{}ive conf\mbox{}igurations were generated.  For each conf\mbox{}iguration, 500 SD gauge f\mbox{}ixings were performed, starting from different randomly chosen points. This
procedure gives an idea on the number of Gribov copies for each conf\mbox{}iguration, and def\mbox{}ines a candidate for the global maximum of (\ref{f}). As expected, the number of Gribov copies seems to increase with the physical volume of the lattice. 

For each conf\mbox{}iguration, we have compared the global maximum obtained with the different 500 SD ('GMAX' in the f\mbox{}igures) with the maxima computed by the CEASD method ('CEASD' in the f\mbox{}igures), by smeared gauge f\mbox{}ixing ('SMGF' in the f\mbox{}igures) and by a single stee\-pest descent applied to the conf\mbox{}iguration ('SD' in the f\mbox{}igures).

\begin{figure}[p]
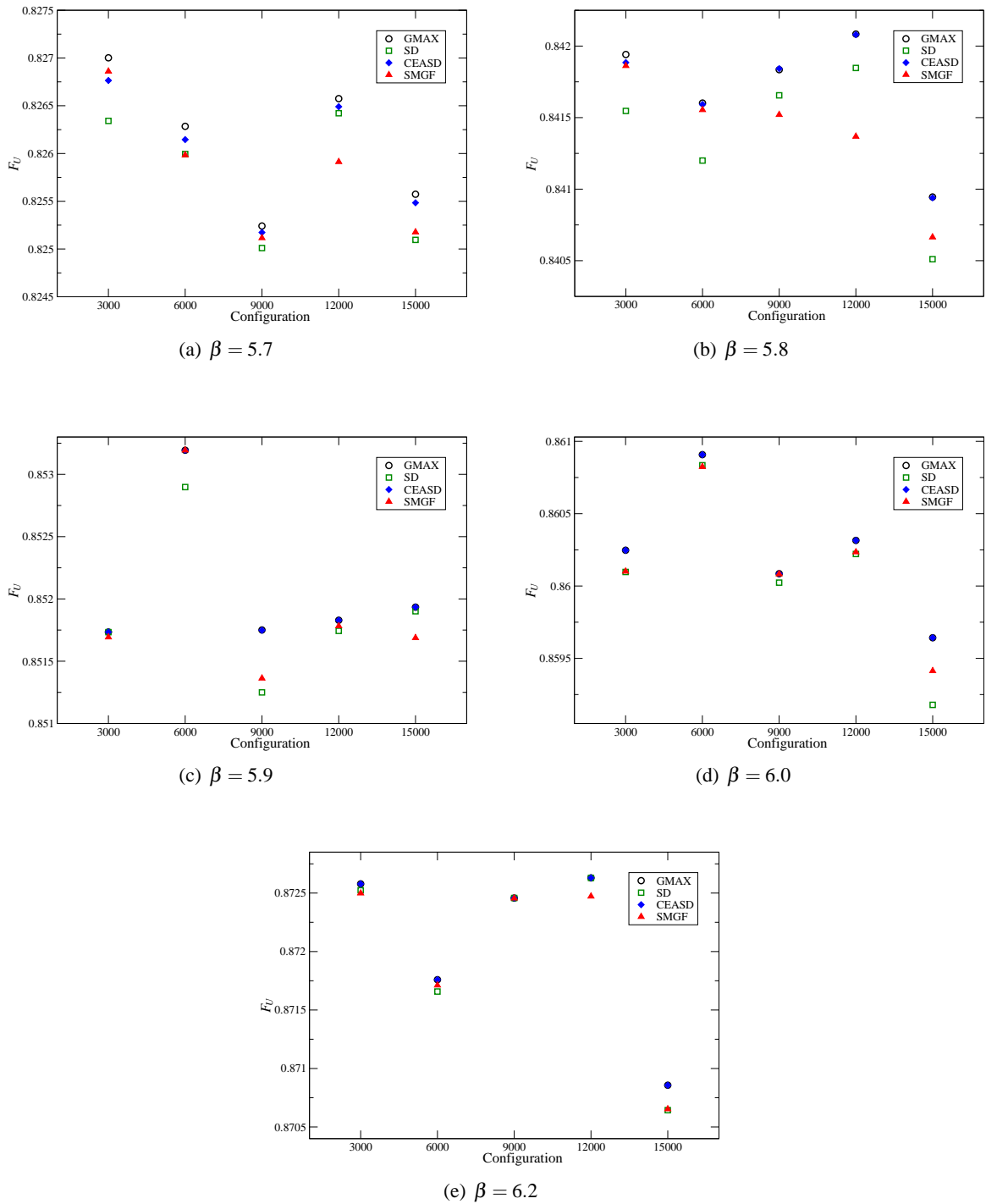

\psfrag{EIXOX}{\hspace*{-0.7cm}{\huge Conf\mbox{}iguration}  }
\psfrag{EIXOY}{{\huge $F_U$}}
  \subfigure[$\beta=5.7$]{
  \begin{minipage}[b]{0.45\textwidth}
    \centering
    \includegraphics[origin=c, angle=0, width=\halffigwidth]{figures/gauge/L16_B57.eps}
  \end{minipage} } \hfill
  \subfigure[ $\beta=5.8$ ]{
  \begin{minipage}[b]{0.45\textwidth}
    \centering
    \includegraphics[origin=c, angle=0,width=\halffigwidth]{figures/gauge/L16_B58.eps}
  \end{minipage} }\vspace*{0.8cm}
  \subfigure[ $\beta=5.9$]{
  \begin{minipage}[b]{0.45\textwidth}
    \centering
    \includegraphics[origin=c, angle=0,width=\halffigwidth]{figures/gauge/L16_B59.eps}
  \end{minipage} } \hfill
  \subfigure[ $\beta=6.0$]{
  \begin{minipage}[b]{0.45\textwidth}
    \centering
    \includegraphics[origin=c, angle=0,width=\halffigwidth]{figures/gauge/L16_B60.eps}
  \end{minipage} }\vspace*{0.8cm}
  \subfigure[$\beta=6.2$]{
  \begin{minipage}[b]{\textwidth}
    \centering
    \includegraphics[origin=c, angle=0,width=\halffigwidth]{figures/gauge/L16_B62.eps}
  \end{minipage} }\vspace*{0.8cm} 
  \caption{$F_U$ maxima found by the different gauge f\mbox{}ixing methods considered.}
\label{gfmeth}
\end{figure}

The various plots in f\mbox{}igure \ref{gfmeth} 
show that the smeared gauge f\mbox{}ixing only once was able to
identify correctly the global maximum ($\beta=5.9$ conf\mbox{}iguration nr. 6000). On the other hand, the CEASD method was successful in all 
$\beta=6.0$ and $\beta=6.2$ conf\mbox{}igurations, and in 4 of the $\beta=5.9$ conf\mbox{}igurations. Curiously, the CEASD method found a larger $F_U$ than the one found by 500 SD for one of the $\beta=5.8$ conf\mbox{}igurations. In what concerns the computation of the absolute maximum of 
$F_U [g]$, the CEASD method seems to be superior, but it is very computationally demanding. Indeed, it requires $\sim$ 30 hours in a single Pentium IV,
to be compared with $\sim 3$ hours for the smeared gauge f\mbox{}ixing method or $\sim 30$ minutes for the steepest descent.

\clearpage

\section{Gribov copies and the gluon and ghost propagators}

The investigation of the Gribov copies effects on the gluon and ghost propagators has been an active research f\mbox{}ield for some time --- see for example, \cite{cucc97,stern05,bbmpm05,bbbimm07} and references there in. Although in a recent work \cite{zwan03} it has been argued that the different maxima of $F_U$ do not change the inf\mbox{}inite volume expectation values, this is not necessarily true on a f\mbox{}inite lattice. 

In this section, the effects of Gribov copies in the gluon \cite{npb2004} and ghost propagators are investigated using 302 conf\mbox{}igurations for a $12^4$ lattice with $\beta=5.8$. For each conf\mbox{}iguration, 500 SD, starting from different randomly chosen points,  were performed and saved the gauge conf\mbox{}igurations associated to the largest maximum of $F_U$, the smallest maximum of $F_U$, and a few more random gauge conf\mbox{}igurations associated to intermediate values of $F_U$ maxima. Then, the propagators computed have been labelled by 
 $\langle F_U \rangle$, the mean value of the corresponding $F_U$ associated to each gauge f\mbox{}ixed conf\mbox{}iguration. 

In this work, we compute the gluon propagator $D(q^2)$ using the following def\mbox{}initions (see \cite{npb2004}  for more details)
\begin{equation}
  D( q^2 ) ~ = ~ \frac{2}{(N^2_c-1)(N_d-1) V} \sum\limits_{\mu} ~ 
                 \langle ~ \mbox{Tr} \left[  A_\mu (\hat{q}) \, 
                                           A_\mu ( -\hat{q} ) \right] ~ \rangle
      \, ,
      \hspace{0.3cm} q \ne 0 \, ,
 \label{Dq2}
\end{equation}
and
\begin{equation}
  D( q^2=0 ) ~ = ~     \frac{2}{(N^2_c-1) N_d V} \sum\limits_{\mu} ~ 
                 \langle ~ \mbox{Tr} \left[  A_\mu (\hat{q}) \, 
                                           A_\mu ( -\hat{q} ) \right] ~ \rangle  \,  ,
    \label{Dq20}
\end{equation}
where
\begin{equation}
   A_\mu ( \hat{q} )  =  \sum \limits_x ~ 
         \exp[-i\hat{q}\left(x + a \hat{e}_\mu/2\right)] ~
         A_\mu (x + a \hat{e}_\mu / 2 ) ,
\end{equation}
and 
\begin{equation}
A_\mu (x + a \hat{e}_\mu / 2) ~ = ~
   \frac{1}{2 i g_0} \Big[ U_\mu (x)  -  U^\dagger_\mu (x) \Big] ~ - ~
   \frac{1}{6 i g_0} \mbox{Tr}
         \Big[ U_\mu (x)  -  U^\dagger_\mu (x) \Big] + \mathcal{O}(a^2).
\end{equation}

On the lattice, the discrete momenta available are given by 
\begin{equation}
  \hat{q}_\mu ~ = ~ \frac{2 \pi n_\mu}{a L_\mu} \, ,
  \hspace{0.5cm}  n_\mu ~ = ~ 0, \, 1, \, \dots \, L_\mu/2 \, ,
\end{equation}
and
\begin{equation}
  q_\mu \, = \, \frac{2}{a} ~ \sin \Big( \frac{\hat{q}_\mu a}{2} \Big) \, .
\end{equation}

The ghost propagator was computed using a plane wave source \cite{cucc97},
\begin{equation}
G^{ab}(\hat{q})=\frac{1}{V}\Big\langle \sum_{x,y}(M^{-1})^{ab}_{xy} \exp[i\hat{q}\cdot(x-y)]\Big\rangle
\label{ghost_ideal}
\end{equation}
and the ghost scalar function computed from
\begin{equation}
G(q^2)=\frac{1}{N_{c}^{2}-1}\sum_{a}G^{aa}(\hat{q}).
\end{equation}

\begin{figure}[!h]    
  \psfrag{EIXOX}{{\huge $\langle F_U \rangle$}}
  \psfrag{EIXOY}{{\huge $D(p^2)$}}
  \subfigure[$n_{\mu}=(0,0,0,0)$]{
  \begin{minipage}[b]{0.45\textwidth}
    \centering
    \includegraphics[origin=c,angle=-90,scale=0.3]{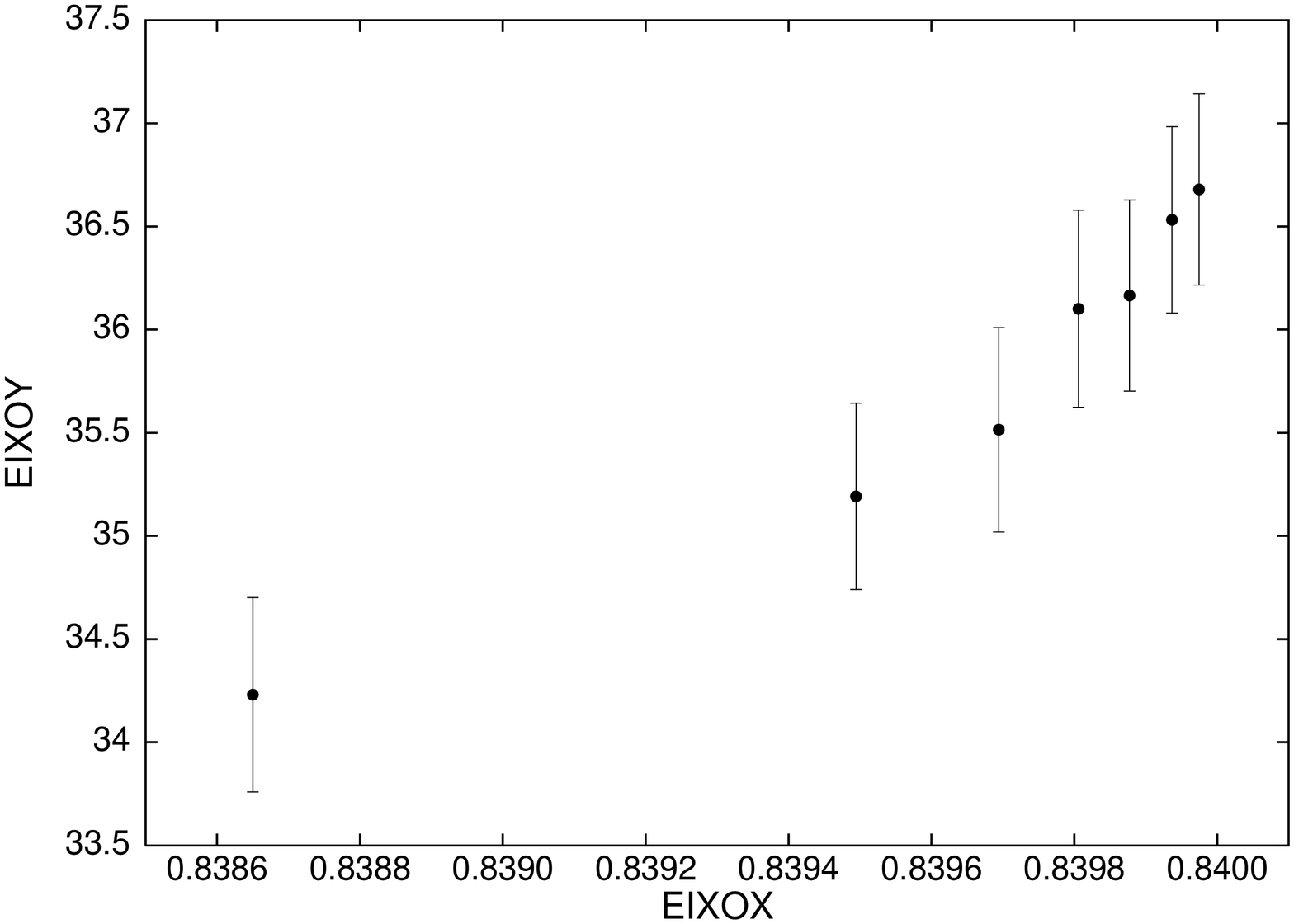}\vspace*{-0.8cm}
  \end{minipage} } \hfill
  \subfigure[$n_{\mu}=(1,0,0,0)$]{
  \begin{minipage}[b]{0.45\textwidth}
    \centering
    \includegraphics[origin=c,angle=-90,scale=0.3]{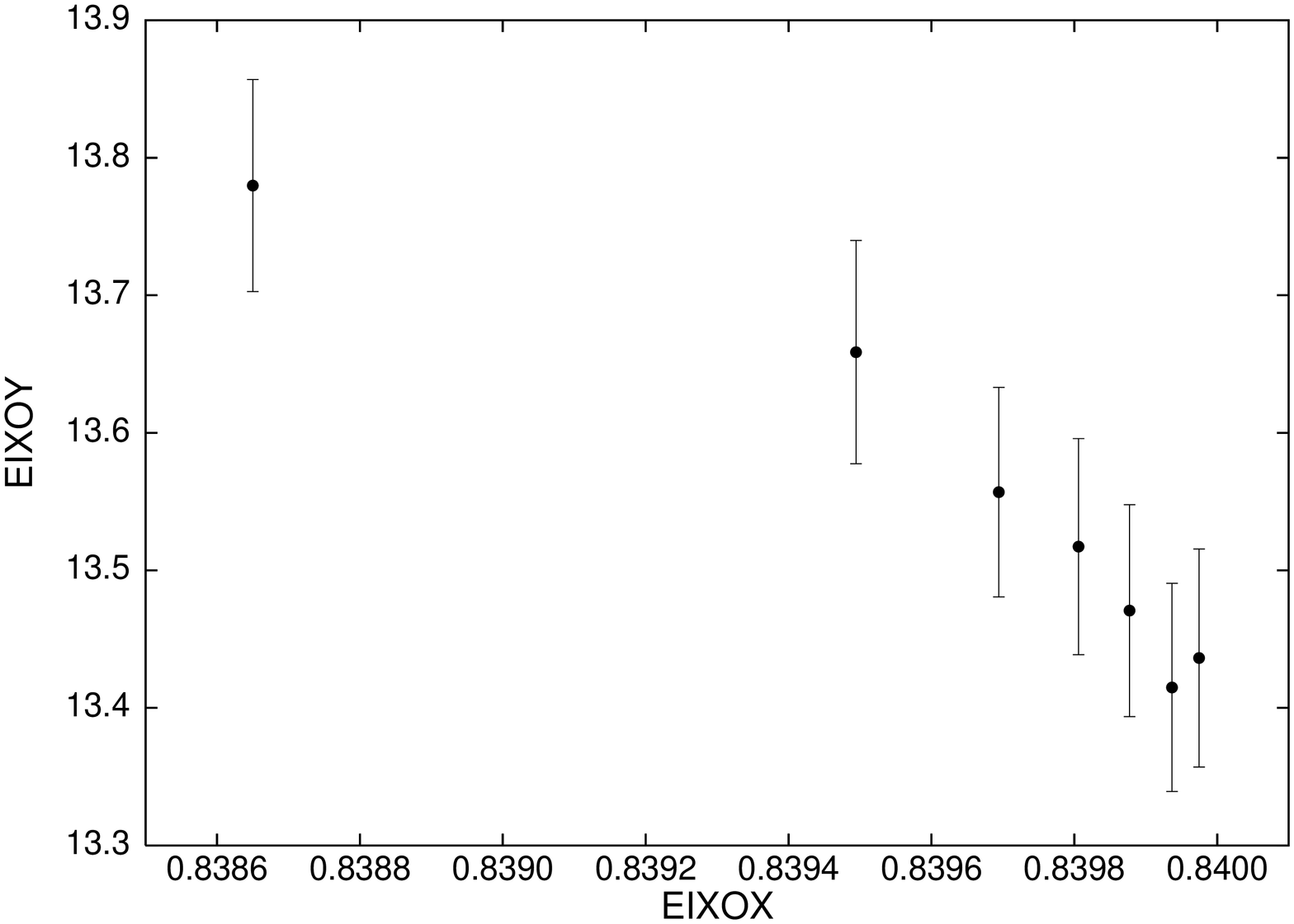}\vspace*{-0.8cm}
  \end{minipage} }\vspace*{0.8cm}
  \subfigure[$n_{\mu}=(1,1,0,0)$]{
  \begin{minipage}[b]{0.45\textwidth}
    \centering
    \includegraphics[origin=c,angle=-90,scale=0.3]{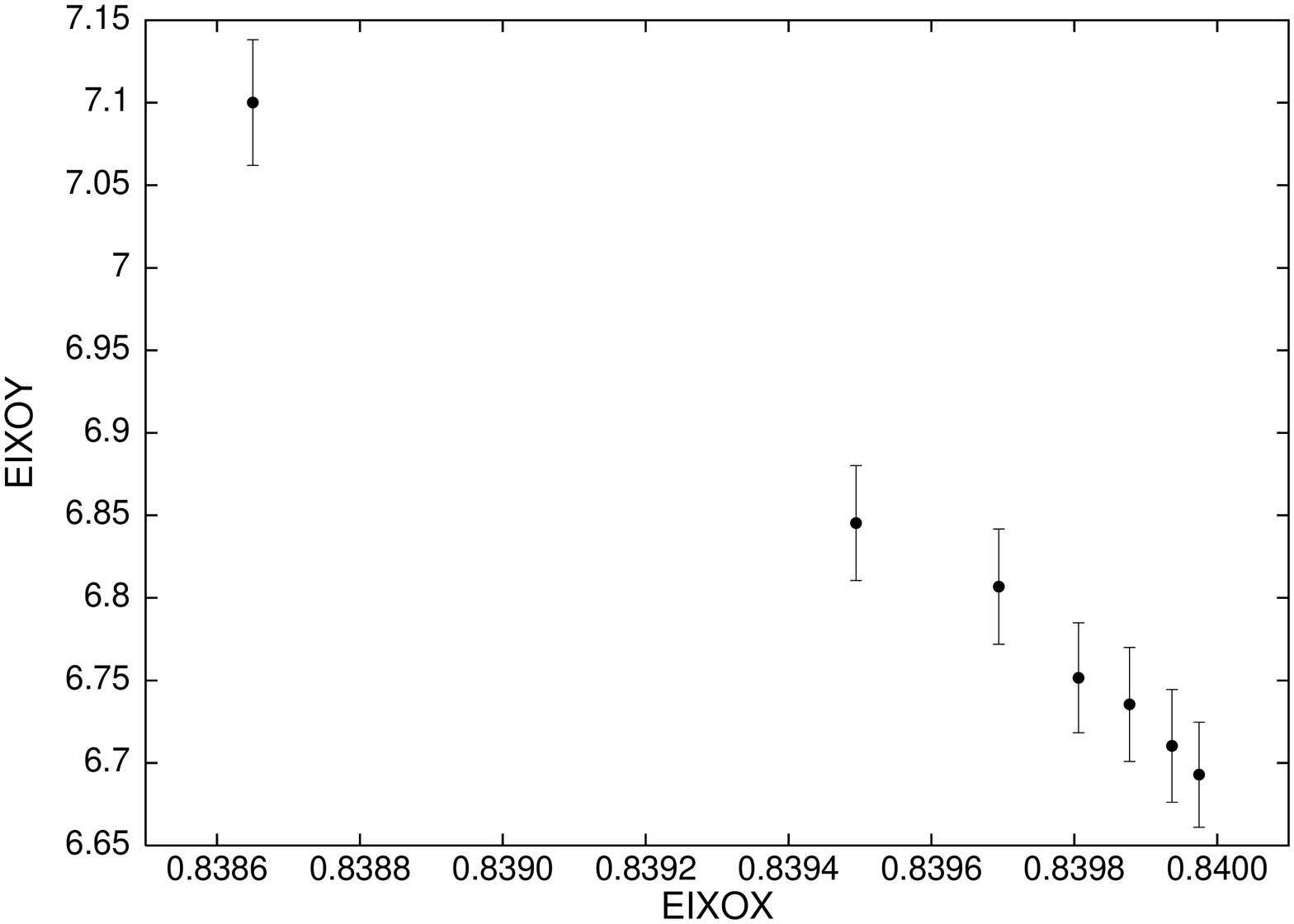}\vspace*{-0.8cm}
  \end{minipage} } \hfill
  \subfigure[$n_{\mu}=(1,1,1,0)$]{
  \begin{minipage}[b]{0.45\textwidth}
    \centering
    \includegraphics[origin=c,angle=-90,scale=0.3]{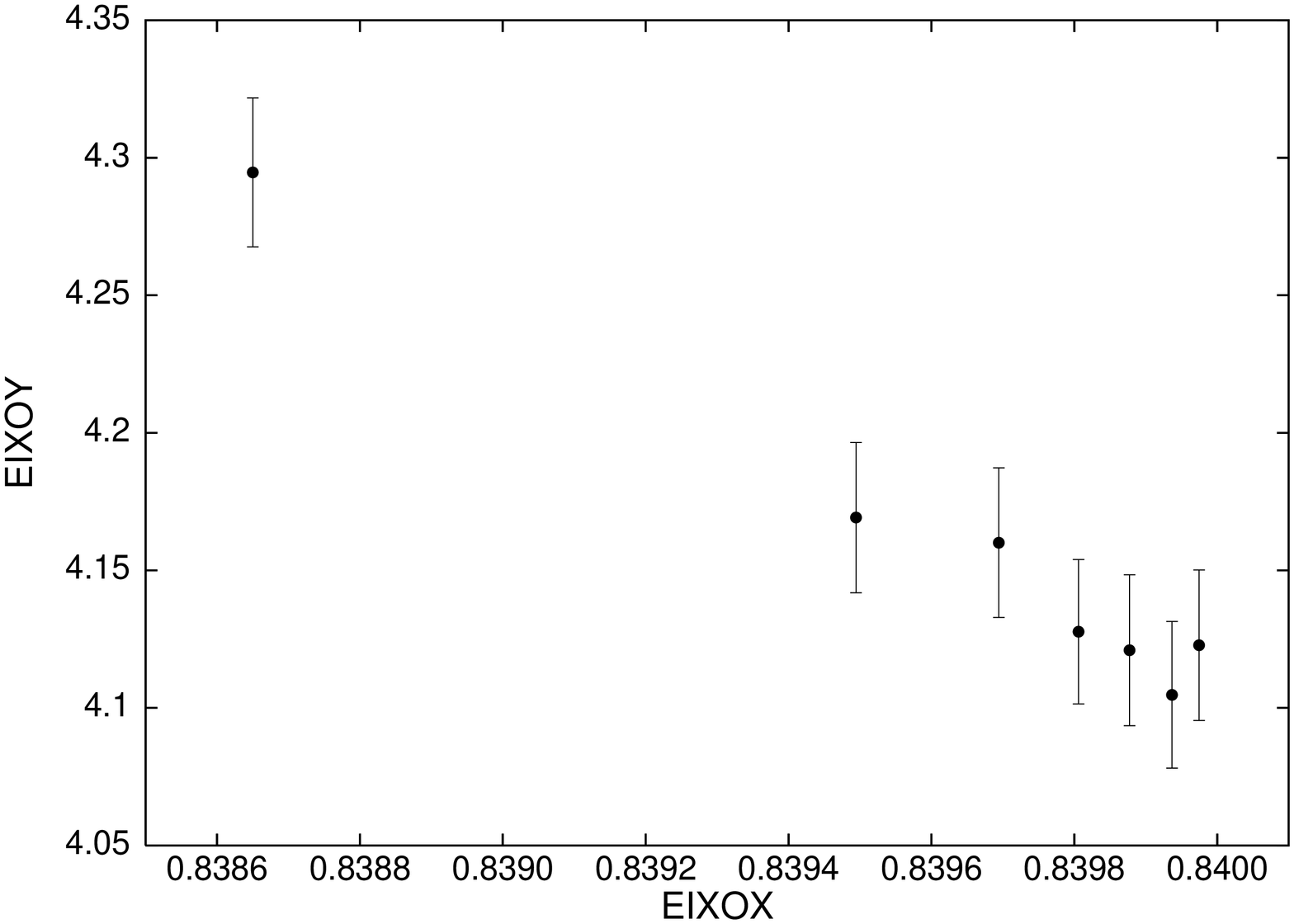}\vspace*{-0.8cm}
  \end{minipage} }\vspace*{0.8cm}
  \subfigure[$n_{\mu}=(1,1,1,1)$]{
  \begin{minipage}[b]{0.45\textwidth}
    \centering
    \includegraphics[origin=c,angle=-90,scale=0.3]{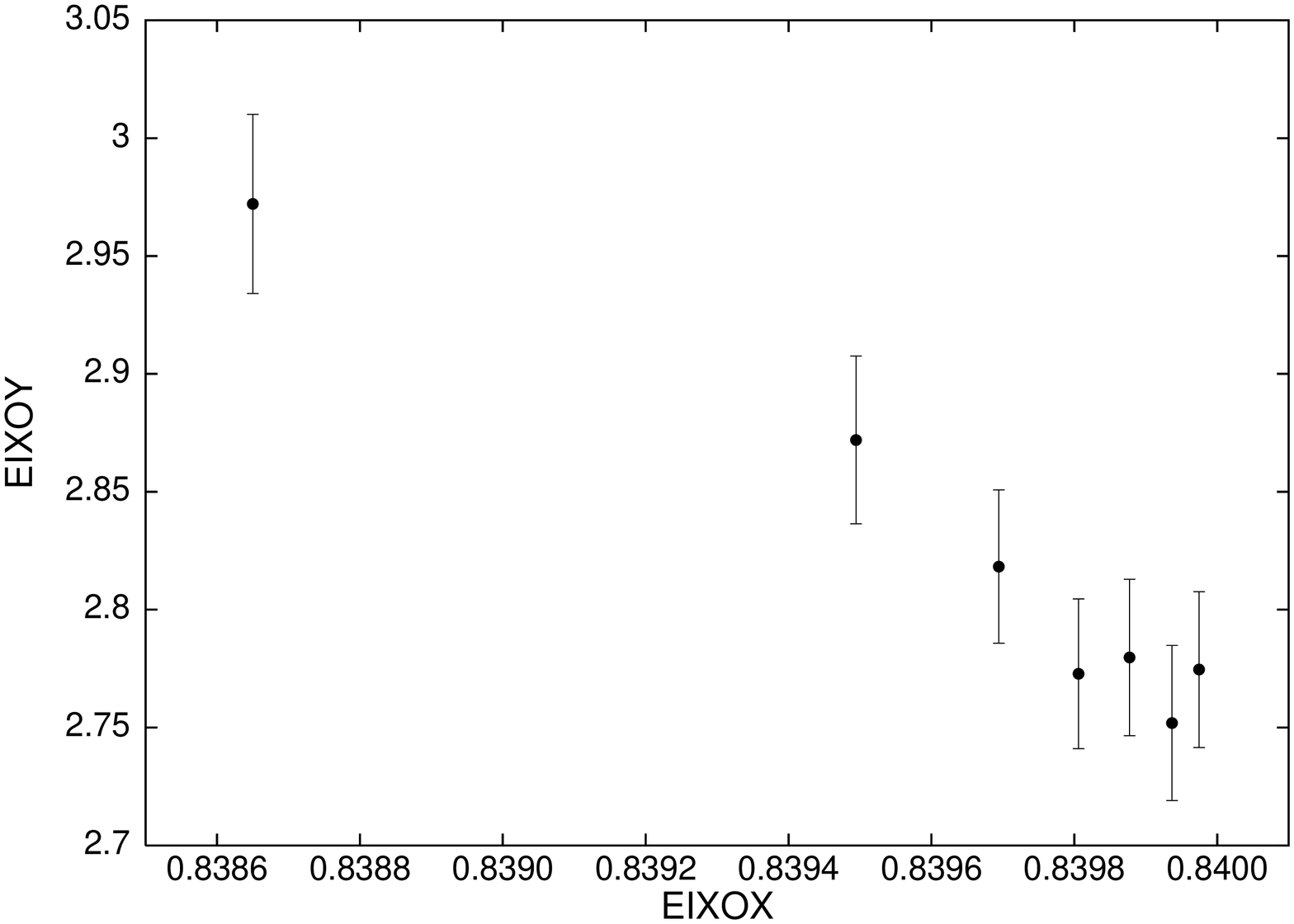}\vspace*{-0.8cm}
  \end{minipage} } \hfill
  \subfigure[$n_{\mu}=(6,6,6,6)$]{
  \begin{minipage}[b]{0.45\textwidth}
    \centering
    \includegraphics[origin=c,angle=-90,scale=0.3]{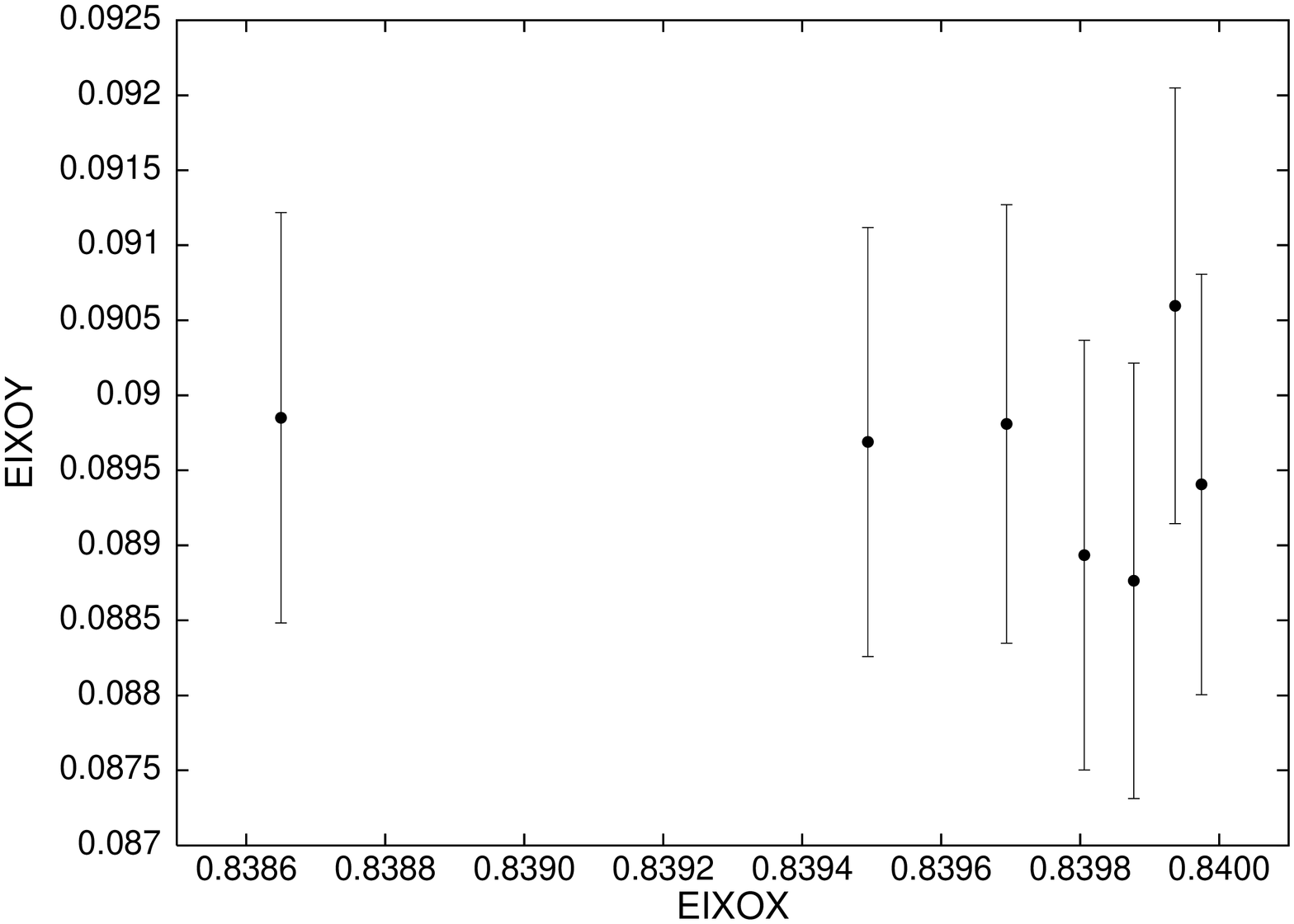}\vspace*{-0.8cm}
  \end{minipage} }\vspace*{0.8cm}
  \caption{Gluon propagator \textit{versus} $\langle F_U \rangle$. }
\label{gluon_raw_F}
\end{figure}

\begin{figure}[!h]
  \psfrag{EIXOX}{{\huge $\langle F_U \rangle$}}
  \psfrag{EIXOY}{{\huge $G(p^2)$}}
  \subfigure[$n_{\mu}=(1,0,0,0)$]{
  \begin{minipage}[b]{0.45\textwidth}
    \centering
    \includegraphics[origin=c,angle=-90,scale=0.3]{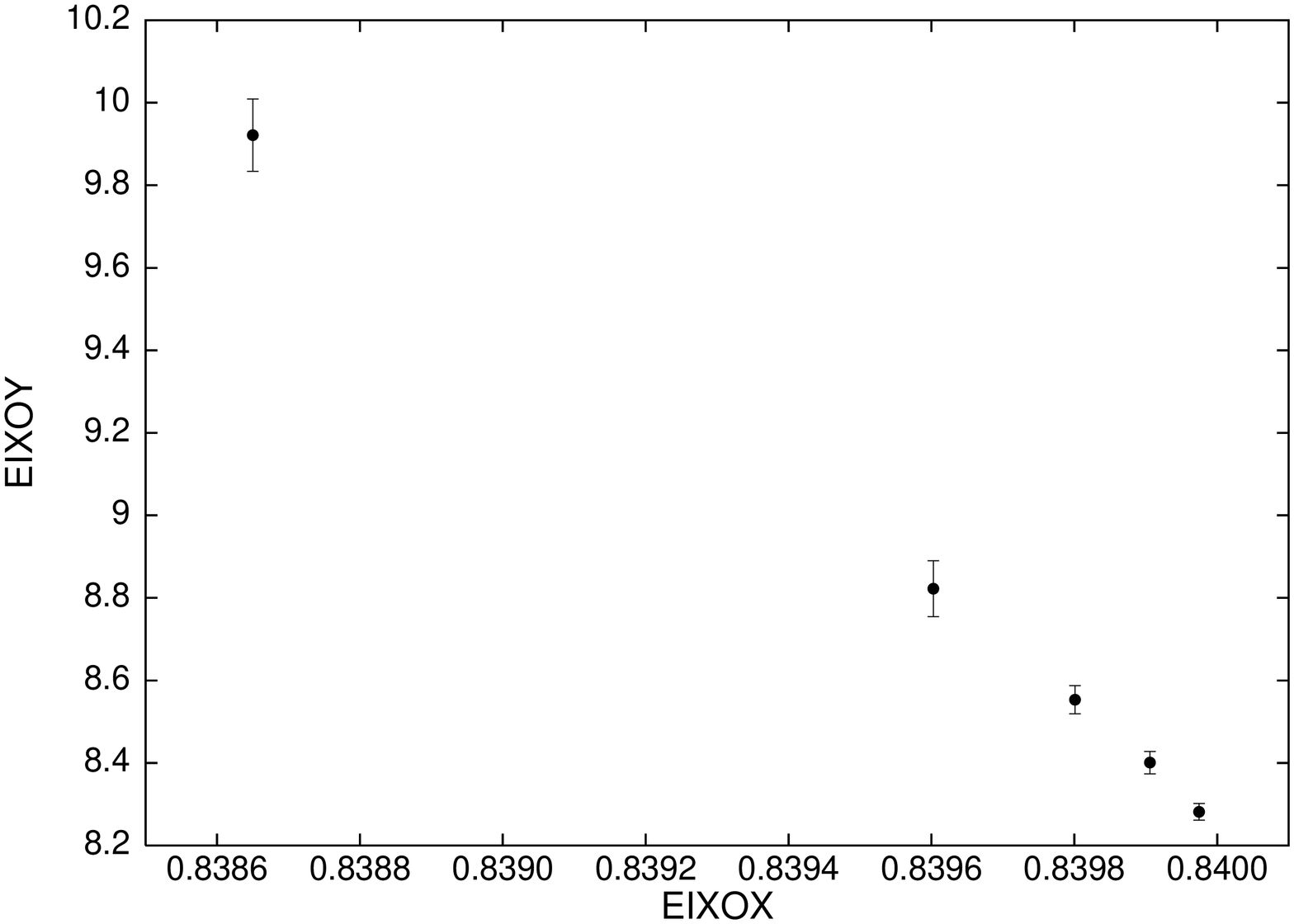}\vspace*{-0.8cm}
  \end{minipage} } \hfill
  \subfigure[$n_{\mu}=(2,0,0,0)$]{
  \begin{minipage}[b]{0.45\textwidth}
    \centering
    \includegraphics[origin=c,angle=-90,scale=0.3]{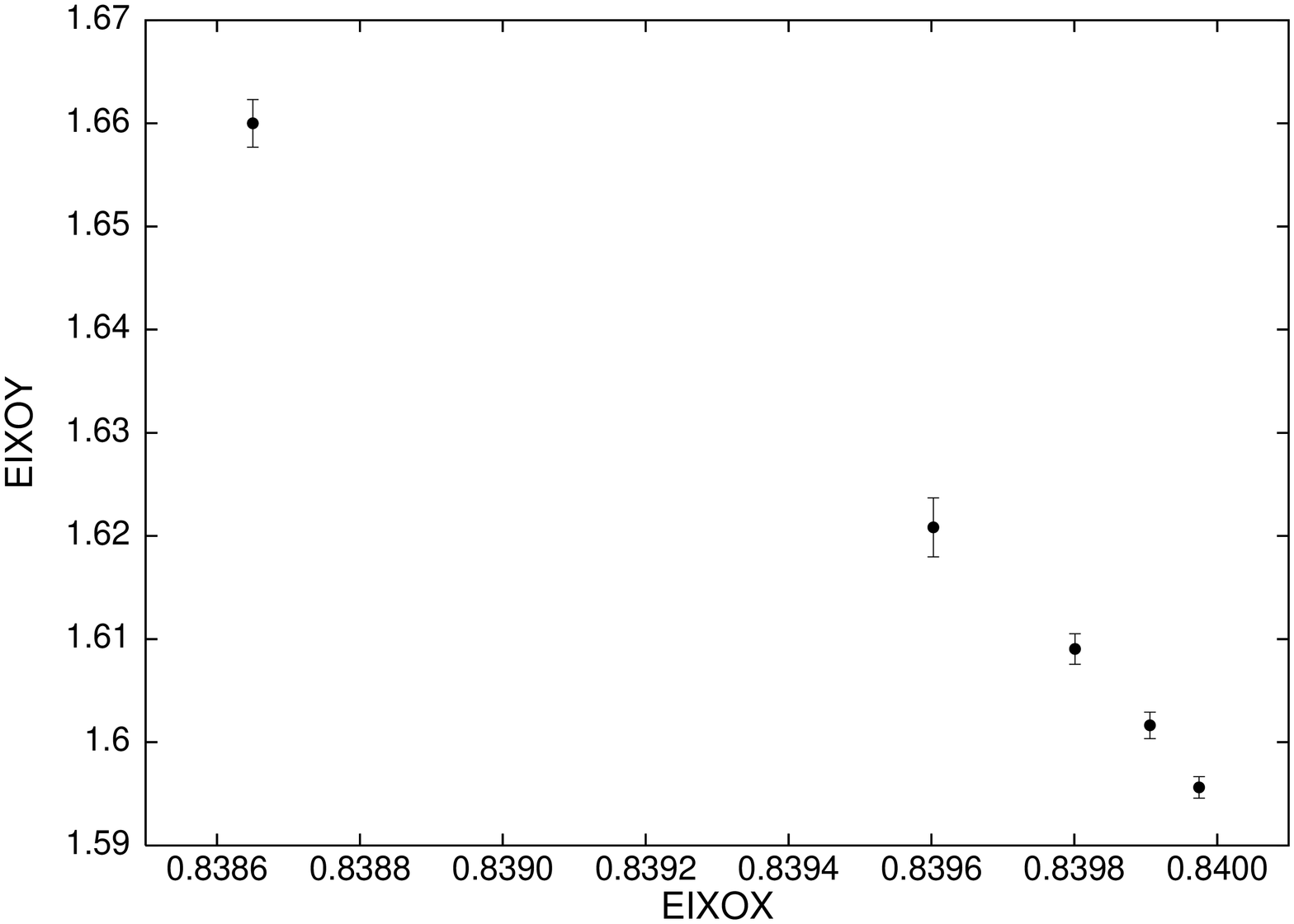}\vspace*{-0.8cm}
  \end{minipage} }\vspace*{0.8cm}
  \subfigure[$n_{\mu}=(1,1,0,0)$]{
  \begin{minipage}[b]{0.45\textwidth}
    \centering
    \includegraphics[origin=c,angle=-90,scale=0.3]{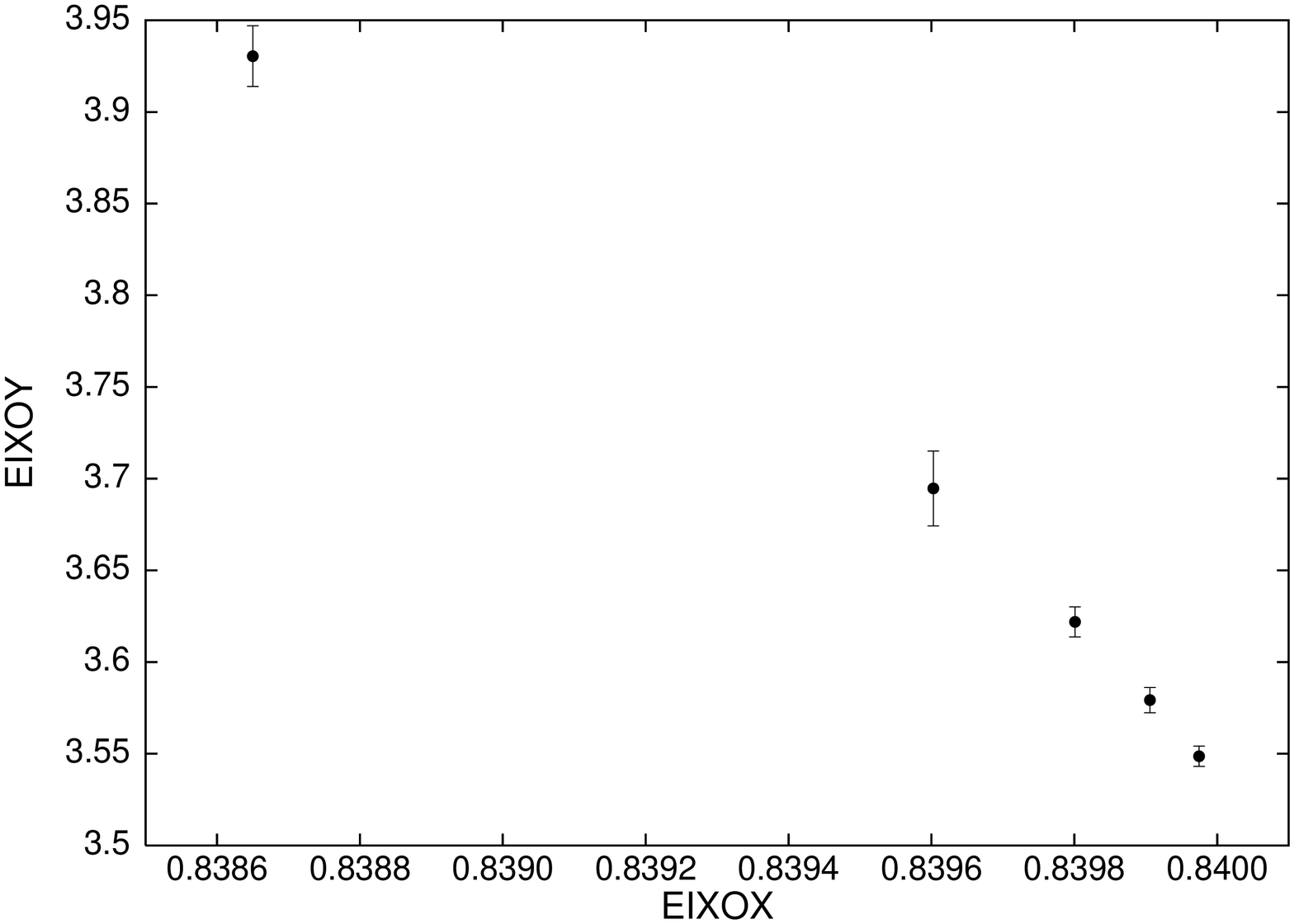}\vspace*{-0.8cm}
  \end{minipage} } \hfill
  \subfigure[$n_{\mu}=(1,1,1,0)$]{
  \begin{minipage}[b]{0.45\textwidth}
    \centering
    \includegraphics[origin=c,angle=-90,scale=0.3]{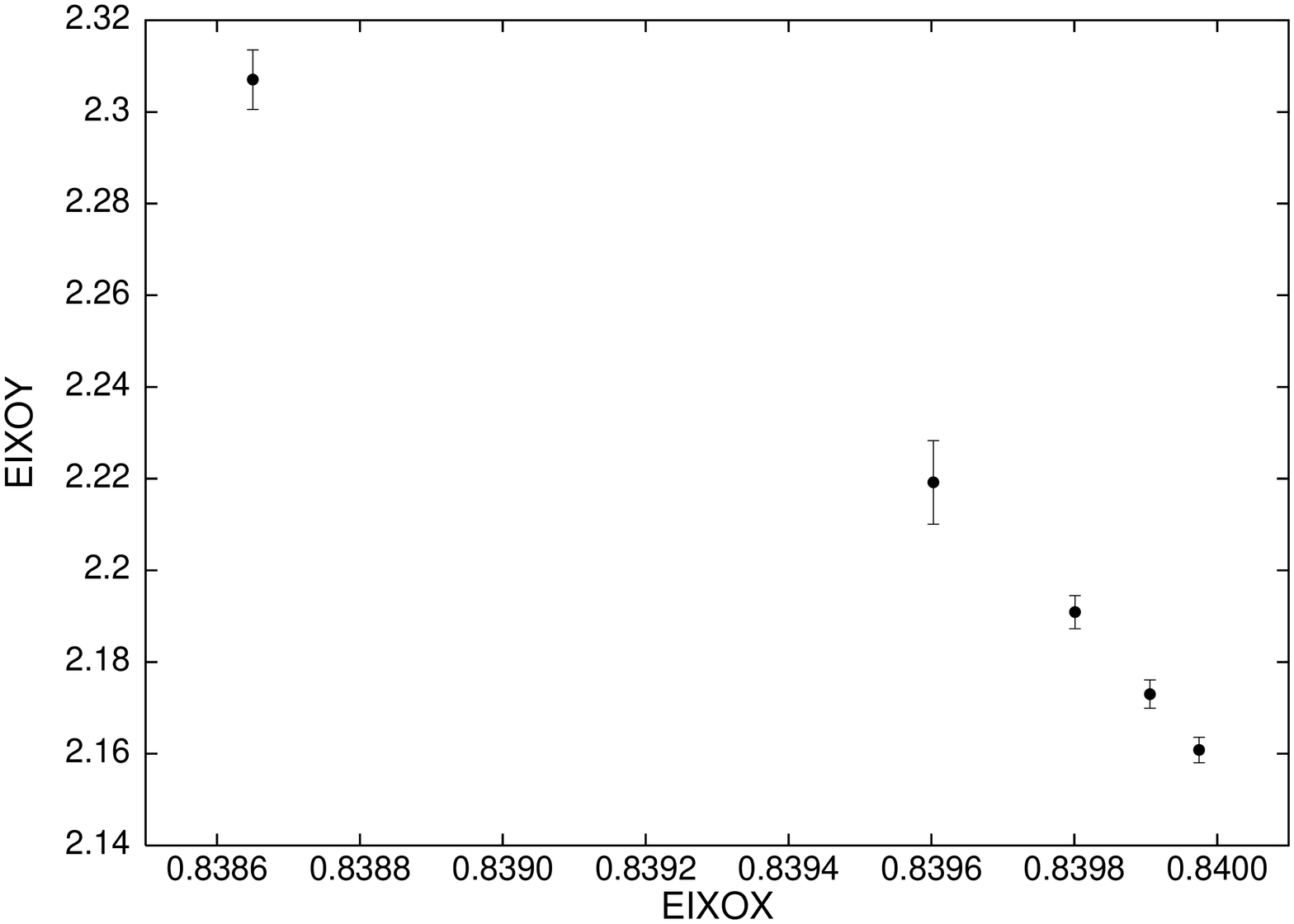}\vspace*{-0.8cm}
  \end{minipage} }\vspace*{0.8cm}
  \subfigure[$n_{\mu}=(1,1,1,1)$]{
  \begin{minipage}[b]{0.45\textwidth}
    \centering
    \includegraphics[origin=c,angle=-90,scale=0.3]{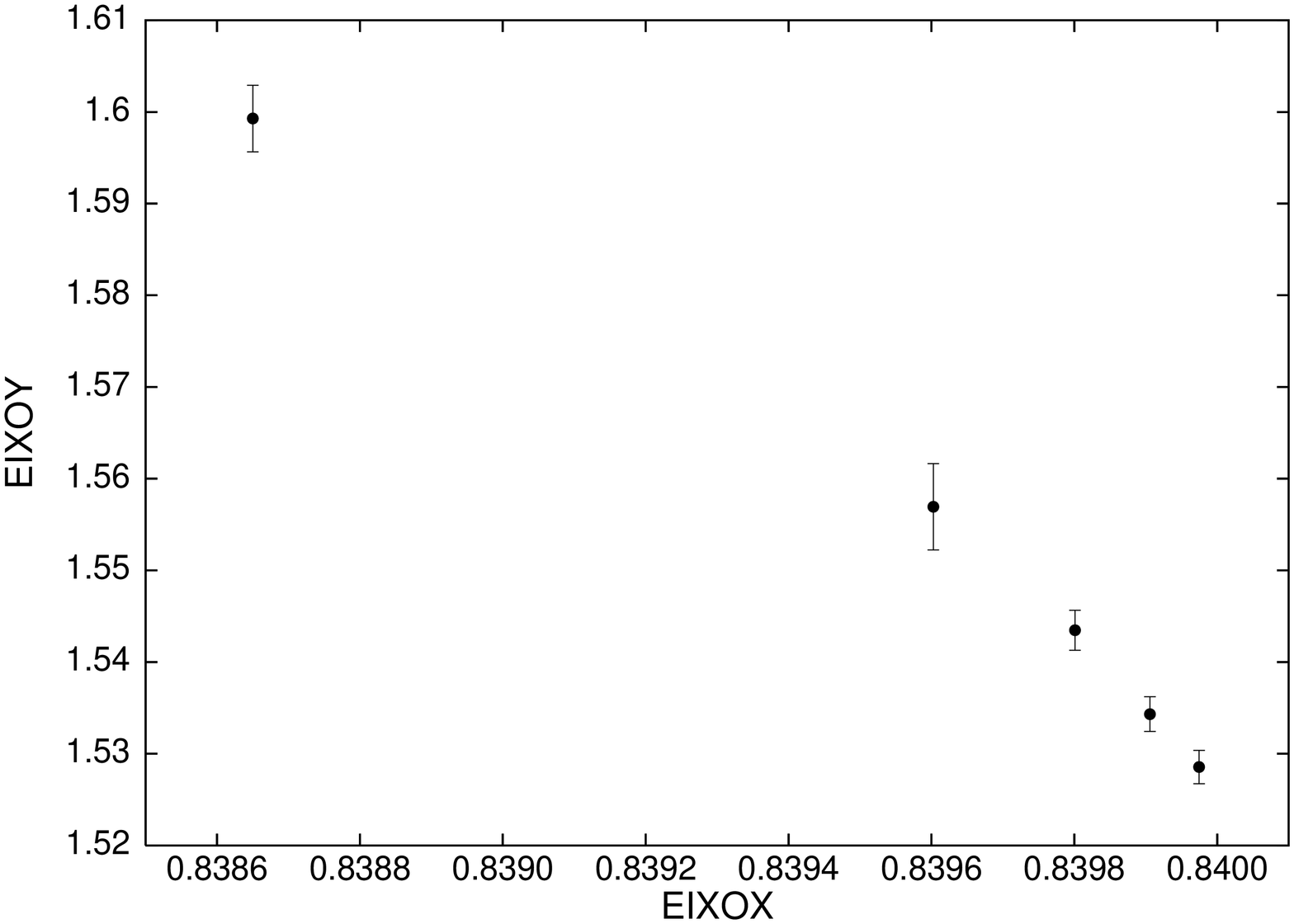}\vspace*{-0.8cm}
  \end{minipage} } \hfill
  \subfigure[$n_{\mu}=(6,6,6,6)$]{
  \begin{minipage}[b]{0.45\textwidth}
    \centering
    \includegraphics[origin=c,angle=-90,scale=0.3]{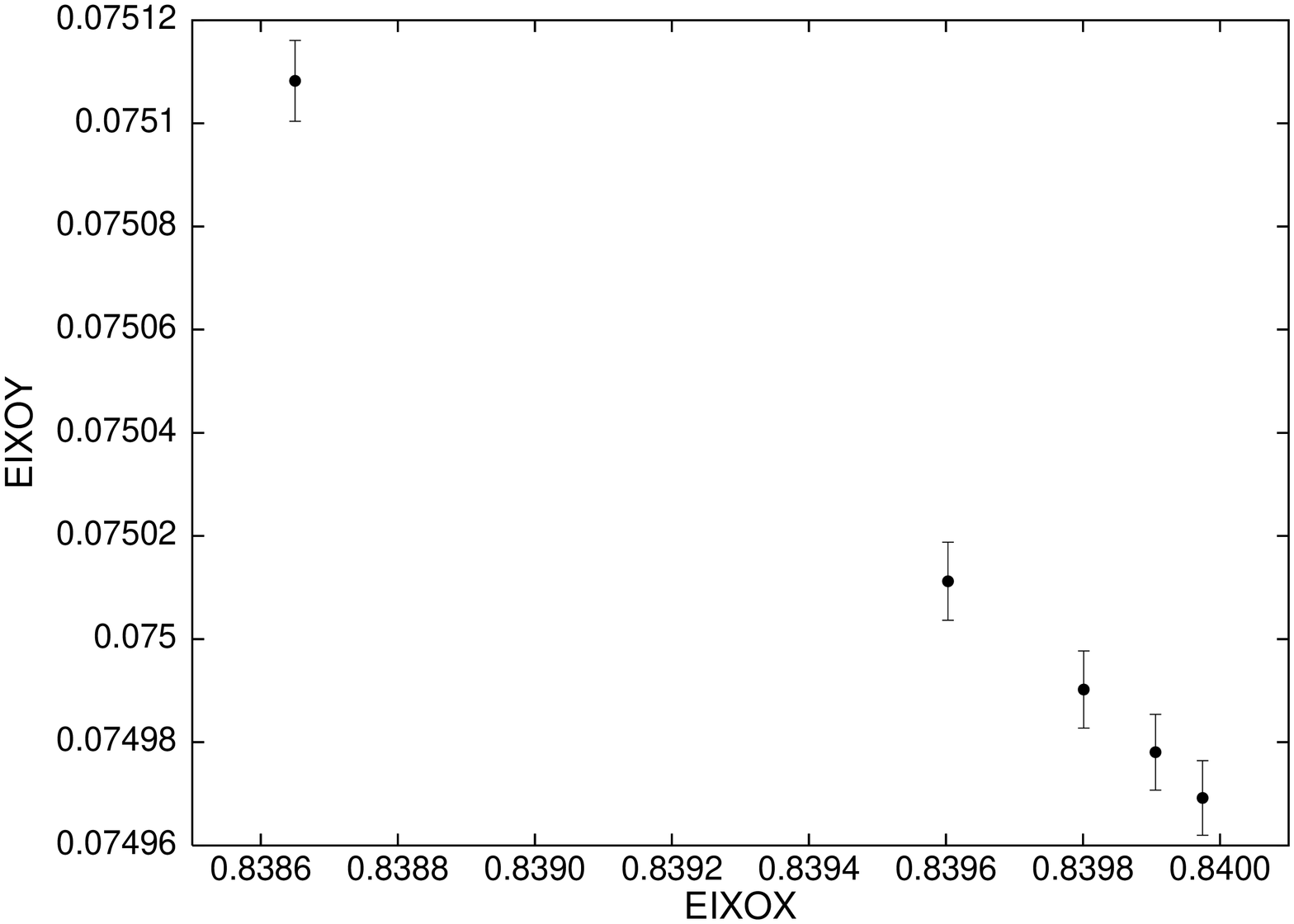}\vspace*{-0.8cm}
  \end{minipage} }\vspace*{0.8cm}
  \caption{Ghost propagator \textit{versus} $\langle F_U \rangle$. }
\label{raw_gh_F}
\end{figure}

In f\mbox{}igure \ref{gluon_raw_F} we report the dependence of the gluon propagator in $\langle F_U \rangle$. The data shows that $D(0)$ increases as $\langle F_U \rangle$ increases. On the other hand,  the low non-zero momenta $D(q^2)$ seem to be decreasing as $\langle F_U \rangle$ increases. The high momenta region seems to be blind to the choice of copy. 
\clearpage
These results show that Gribov copies are important only for the infrared regime. In what concerns the ghost propagator, see f\mbox{}igure \ref{raw_gh_F}, their values decrease as $\langle F_U \rangle$ increases. Once more, the effect is bigger for low momenta (a 20\% effect for the lowest momentum,  reduced to 0.2\% for the highest momentum considered).

\section{Conclusions} 

For the lattice Landau gauge, we have compared two methods of gauge f\mbox{}ixing tunned to approach the global maximum of $F_U[g]$. For a $16^4$ lattice, the CEASD method seems to perform better than smeared gauge f\mbox{}ixing. Unfortunately, CEASD is quite computational demanding, not only in what concerns memory but also in CPU time.

In the second part of this paper, we investigated the effect of Gribov copies in the gluon and ghost propagators. Our data shows a clear dependence of gluon and ghost propagators on $\langle F_U \rangle$. Furthermore, the data shows that the changes on the propagators due to the choice of the maximum of $F_U [g]$ are mainly on the infrared region.

\acknowledgments

This work was supported by FCT via grant SFRH/BD/10740/2002, and partially supported by projects POCI/FP/63436/2005 and POCI/FP/63923/2005.

\end{document}